# Twistor diagrams for all tree amplitudes in gauge theory: a helicity-independent formalism

Andrew Hodges
andrew.hodges@wadh.ox.ac.uk
*Wadham College, University of Oxford, Oxford OX1 3PN, United Kingdom*

December 2005

*Abstract:* We give a new formalism for pure gauge-theoretic scattering at tree-amplitude level. We first describe a generalization of the Britto-Cachazo-Feng recursion relation in which a significant restriction is removed. We then use twistor diagrams to express all tree amplitudes in a form independent of helicity. A formal procedure involving anticommuting elements is required. We illustrate the results with specific calculations of interest, up to 8 interacting fields.

**1. Introduction:** An earlier paper (Hodges 2005) showed how the twistor diagram formalism gives a calculus for all tree amplitudes in pure gauge-theoretic scattering. The essential advance lay in showing that the recursion relation of Britto, Cachazo and Feng (2004b) has a natural transcription into twistor diagrams. This earlier paper also mentioned a number of geometric features of the resulting structure suggesting scope for further generalization and simplification.

We now briefly describe how these geometric features can be exploited to achieve a more powerful and unified formalism. Essentially, our goal is to treat each gauge field as a unity, without splitting it into helicity parts. In the description of $n$-field processes, this will allow us to combine $2^n$ cases within a unified scheme. Our method for doing this rests on extending the BCF relation so as to allow a helicity combination hitherto forbidden. This extension requires a formal procedure for 'super-expanding' the boundary-lines of twistor diagrams into finite sums of terms, with certain combinatorial coefficients, and we shall express this formal procedure by a rule involving anticommuting elements. At this stage, we shall regard the procedure simply as a concise means of specifying the finite set of terms, so that the diagrams continue to represent exactly the same kinds of twistor-space contour integrals as before. But it may in future be of interest to see whether a link can be found with topological strings in a 'supertwistor space', as the form of the emergent diagrams seems to indicate. Alternatively, following the strategy of the Penrose 'twistor programme', one might seek to express this procedure through new developments in twistor geometry rather than by the addition of formal anticommuting variables.

**2. Expansion of twistor propagators.**

In what follows we build on the notation and results of (Hodges 2005). Consider first the basic element of twistor diagrams: the *boundary* element written

$$W \sim\!\sim\!\sim\!\sim Z$$

and defining a boundary for the contour on the subspace $W_\alpha Z^\alpha = k$

Such an element may be composed with a twistor function $f(Z)$ of homogeneity degree (–4) to effect a twistor transform. But suppose it is composed with a function of degree (–3). Using integration by parts, we find that

$$f_{-3} \!\!\!\succ\!\!\bullet\!\!\sim\!\sim\!\sim = k \quad f_{-3}\!\!\!\succ\!\!\bullet\!\!\!-\!\!\!-\!\!\!-$$

There are similar results for functions of degree –2, –1, 0, which can then be written as an *expansion*:

$$\succ\!\!\bullet\!\!\sim\!\sim = \succ\!\!\bullet\!\!\sim\!\sim + k \succ\!\!\bullet\!\!-\!\!- \\ + \frac{k^2}{2!} \succ\!\!\bullet\!\!=\!\!= + \frac{k^3}{3!} \succ\!\!\bullet\!\!\equiv\!\!\equiv \quad (1) \\ + \frac{k^4}{4!} \succ\!\!\bullet\!\!\equiv\!\!\equiv$$

where each term on the right-hand side contributes only in a sector of particular homogeneity degree. Similar expansions can be carried out in more general diagrammatic settings.

In (Hodges 2005) we defined the 'twistor quilt' corresponding to a twistor diagram. A quilt is obtained by replacing all the lines by boundary-lines. Thus it simply represents an element in (relative) homology. We commented on the fact that for all the diagrams obtained by the BCF rule, one may replace the diagram by the quilt, and the resulting integral will be unchanged apart from factors of $k^4/24$. This comment was based on applying the foregoing 'expansion' argument.

The reason for singling out for attention this 'quilt' feature of the diagrams is that there are notable identities and linear relationships between diagrams. Highly non-trivial identities arise from the fact that amplitudes can be given different representations by using different pivot fields. The 'quilt' property suggests that it should be possible to represent these remarkable identities and relationships as statements about homology alone: *i.e.* as purely geometric structure. Investigation shows, however, that this cannot be the case for the diagrams as so far defined, because the truth of the identities depends on the specific choice of positive and negative helicities on the outside, *i.e.* the external homogeneities. The development now to be introduced can be considered as motivated by a desire to remove this unwanted dependence, and so derive a more purely geometric theory.

**3. The super-expansion rule using formal anticommuting elements:**

Super-expanded diagrams will be defined in this section as certain sums of twistor diagrams with certain combinatorial coefficients, prescribed by a rule. To define the rule we enter into a formal procedure which involves supersymmetric elements. The procedure is as follows: we first consider a formal scheme in which the number $k$ on a boundary line from $W$ to $Z$ is replaced by an object $\hat{W}_i \hat{Z}^i$, where the $\hat{W}_i$, $\hat{Z}^i$ are *anticommuting*. The index $i$ runs from 1 to 4. Then we write out the formal expansion of this boundary line following the form of (1). We will call this a 'super-expansion'. The powers of the anticommuting elements in the resulting numerators give rise to combinatorial factors, *e.g.*

$$\left(\hat{W}_i \hat{Z}^i\right)^4 = 4! \, \hat{W}_1 \hat{W}_2 \hat{W}_3 \hat{W}_4 \hat{Z}^1 \hat{Z}^2 \hat{Z}^3 \hat{Z}^4$$

and further combinatorial factors also arise when the numerators loop: e.g. the 4 in:

$$\left(\hat{W}_i \hat{Z}^i\right)^3 \left(\hat{W}_j \hat{X}^j\right) \left(\hat{Y}_k \hat{X}^k\right)^3 \left(\hat{Y}_l \hat{Z}^l\right) = 4 \cdot 3! \, 3! \, \hat{W}_1 \hat{W}_2 \hat{W}_3 \hat{W}_4 \hat{Z}^1 \hat{Z}^2 \hat{Z}^3 \hat{Z}^4 \hat{Y}_1 \hat{Y}_2 \hat{Y}_3 \hat{Y}_4 \hat{X}^1 \hat{X}^2 \hat{X}^3 \hat{X}^4$$

These combinatorial factors give the essential content of the 'super-expansion'. All we need to do is to capture them. We can do this simply by replacing this formal sum with a corresponding sum of twistor diagrams, each with the $k$ in its inhomogeneous boundary or pole, but retaining these coefficients. (An example of such a sum is given in section 5.1 below.) I owe to David Skinner the observation that this replacement procedure is equivalent to defining the boundaries to be given formally by

$$W_\alpha Z^\alpha + \hat{W}_i \hat{Z}^i - k = 0$$

and then formally expanding in powers of $\hat{W}_i \hat{Z}^i$. If anticommuting elements are also associated with the external fields in the natural way, then formal integration over the anticommuting variables will project out just the correct terms.

We introduce some new notation: an *arc* represents a super-expanded boundary as described above. (As the purpose of this formal definition is to express everything in terms of super-expanded boundaries, we do not need notation for an analogous super-expansion of poles.) Now for the inner product of any fields with helicities in the range from $-1$ to $1$ we can simply write:

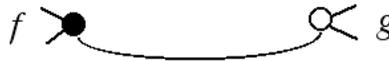

and we also have the double-transform identity:

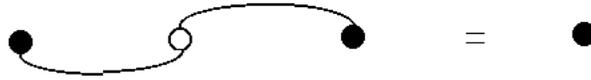

In each case the combinatorial coefficients arising from the powers of $\hat{W}_i \hat{Z}^i$ act simply so as to cancel the unwanted factors in (1). In the next section we shall use the further combinatorial coefficients that arise from a loop of numerators.

**4: BCF recursion and quadruple cuts**

The key material here is the BCF recursion relation. As shown in (Hodges 2005) this can be translated into twistor diagrams as folllows. We assume there are $n$ gauge fields (1, 2, 3,... $x$, $y$... $n$) in cyclic order, and that twistor diagrams are known for all processes of degree less than n. Then the complete amplitude for the $n$ fields is given by the sum over $i$ from $y+1$ to $x-2$ of the two types of diagram:

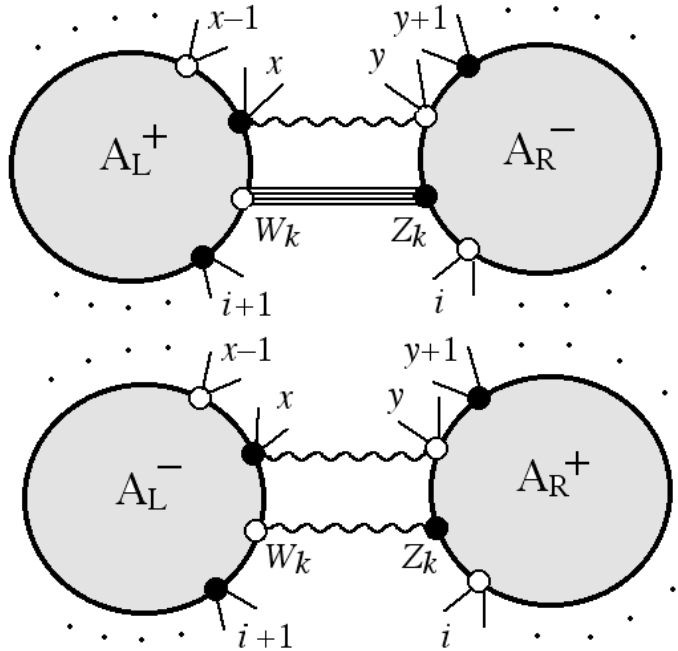

Here $A_L^+$ represents the subamplitude $A(k^+, i+1... x–1, x)$ and similarly for the others. In that earlier paper, we treated only the case of ($x$–, $y$+), *i.e.* the case when both 'bridge-end' or 'pivot' external fields are represented by functions of degree –4. In this we followed the earlier paper of Britto, Cachazo and Feng which used only this case. In the later paper of Britto, Cachazo, Feng and Witten (2005), an extension was made to the cases of pivot pairs ($x$–, $y$–) and ($x$+, $y$+). This corresponds to at least one pivot field being of degree –4. The demonstration of the twistor diagram correspondence goes through unchanged for that extension, since that proof makes no mention of the homogeneity.

However, the BCFW proof does not apply to the remaining case ($x$+, $y$–), corresponding to pivot fields both of degree 0. In terms of 'momentum shifts', characterized by a complex variable $z$, there is a pole at $z = \infty$ which prevents a simple decomposition into partial fractions.

We can fill this gap by combining their work with an observation of Roiban, Spradlin and Volovich (2004). This refers to the *quadruple cut* method of finding 'box function coefficients' and thereby evaluating amplitudes, due to Britto, Cachazo and Feng (2004a). All of this depends upon the special properties of N=4 supersymmetric gauge theory at the one-loop level. However, it must be emphasised from the start that we are here only using the deductions that can be made regarding tree amplitudes, where the N=4 extension cannot make any difference. Thus, we are using the formulas arising from N=4 theory purely as mathematical expressions, without any need to posit N=4 gauge theory as corresponding to anything of physical significance.

Roiban, Spradlin and Volovich (2004) deduce from the infra-red behaviour of N=4 theory at one-loop level that for the tree-level amplitude A, we have:

$$2\,A = \sum_{i=y+1}^{x-2} \quad \text{[box diagram with external lines } x, y, x{-}1, i{+}1, i, y{+}1\text{]}$$

where the diagrams on the right-hand side represent '1-mass' or 'two-mass-hard' box coefficients.

When there is just *one* field at some corner, as happens in all but the '4-mass' coefficients, the resulting expression naturally bifurcates into a sum of two, the splitting being by α-planes and β-planes in complex momentum space. In the case of the '1-mass' and '2-mass-hard' diagrams, we can represent this bifurcation by

$$2\,A = \sum_{i=y+1}^{x-2} \quad \text{[box diagram 1]} \;+\; \text{[box diagram 2]}$$

We now take the case where the two pivotal fields are of opposite helicity, say $(x+\; y-)$. Then the first of these sums is precisely the same, term by term, as the BCF summation using the same pivots.

It follows by subtraction that we have another amplitude formula given by

$$A = \sum_{i=y+1}^{x-2}$$

[diagram: box diagram with external legs $x$, $y$ at top (connected to filled and open vertices), and $x-1$, $i+1$, $i$, $y+1$ at bottom]

and this is exactly what is needed for the missing case in the BCF rule. (This result suggests that every infra-red identity can in fact be separated into two identities.)

Next, we use the idea that the quadruple cut formalism can be immediately translated into twistor diagrams. This follows from the observations that (1) the subamplitude objects at each vertex have twistor diagram translations, by the BCF rule or otherwise, (2) the lines joining the four sub-amplitudes are simply on-shell propagators and so correspond to simple twistor diagram lines, and (3) the bifurcation into two parts also translates into twistor diagrams in the obvious manner.

We further assume that this analysis extends to cases where the two pivot fields are scalars or fermions in the supersymmetric scheme. This is just an aspect of supersymmetric invariance. We may note in support of this assumption that the infra-red analysis depends only on the external momenta, not on helicity. The bifurcation for null external momenta is also independent of external helicities. We refer to this supersymmetric extension as the extended BCFRSV rule.

We can now establish our main result by induction. We actually prove something which is stronger than what we shall need in the end:

**Extended super-expanded diagram recursion rule:** Consider $n$ external fields, where the adjacent fields labelled $x$ and $y$ may be any members of the field supermultiplet, and the others are all of spin 1. Then their complete tree amplitude is given simply by the sum of super-expanded twistor diagrams:

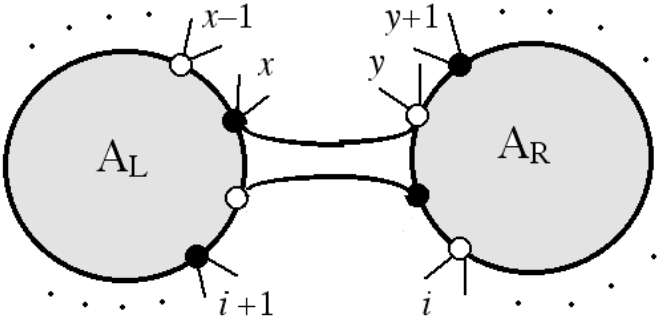

The gauge fields are to be considered as attached without being split into $+$ and $-$ parts.

First, we check that the rule is true for the generation of 4-field amplitudes out of formal 3-field amplitudes: this gives a start to the induction. For this we need the formal 3-amplitudes: these are just:

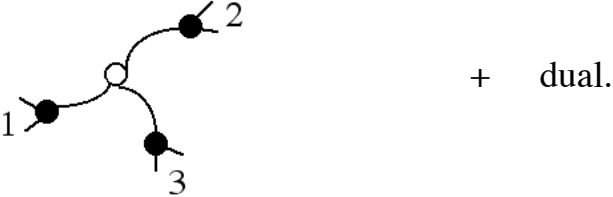     $+$    dual.

For the inductive step, we need only translate from the extended BCFRSV rule to the corresponding twistor diagram. In every case to be considered we find that the on-shell propagators in the quadruple-cut diagram correspond to terms in the super-expansion of the corresponding boundary in the twistor diagram. The inductive hypothesis ensures that the subamplitude diagrams exist and have the requisite properties. There are various cases to consider, depending on the helicities of the pivot fields, but in every case the allowed helicity flows and corresponding subamplitudes correspond exactly to the non-zero possibilities for super-expansion.

In each case also the cyclic product of the formal $\hat{W}_i \hat{Z}^i$ numerators exactly reproduces the numerical factors that in N=4 field theory are contributed by the trace of the super-fields running round the loop.

This completes the demonstration. We now restrict to the case of interest: where the external fields are spin-1 gauge fields of either helicity.

**Super-expanded diagram recursion rule:** Consider $n$ external gauge fields of any helicity. Then their complete tree amplitude is given by the sum of super-expanded twistor diagrams:

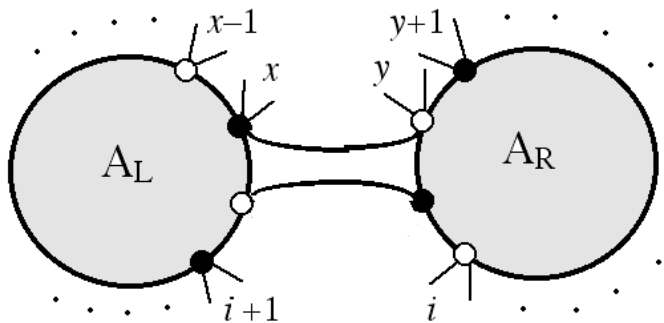

Thus, our claim is that amplitudes are completely specified by super-expanded quilts. There is a potential generalization allowing all the external fields to be fermions or scalars within a SUSY setting, but we shall not examine this further here. Instead, we emphasise that this final result does not depend on N=4 theory. It is purely mathematical and could have been obtained in another way, *e.g.* by analysing the obstruction to the BCFW partial fraction expansion in the excluded case.

**5: Examples of use:**

We will briefly illustrate the general recursive rule by examples.

**5.1:** A(1234) for all helicities:

Take pivots (41) with 4 on a dual twistor and 1 on a twistor. The recursion rule gives the diagram:

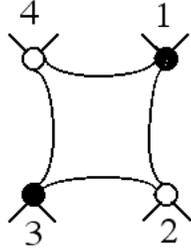

Examination of the 'hard' case A(1+2–3+4–) will illustrate the economy of representation that has been achieved. Super-expansion of the boundaries gives:

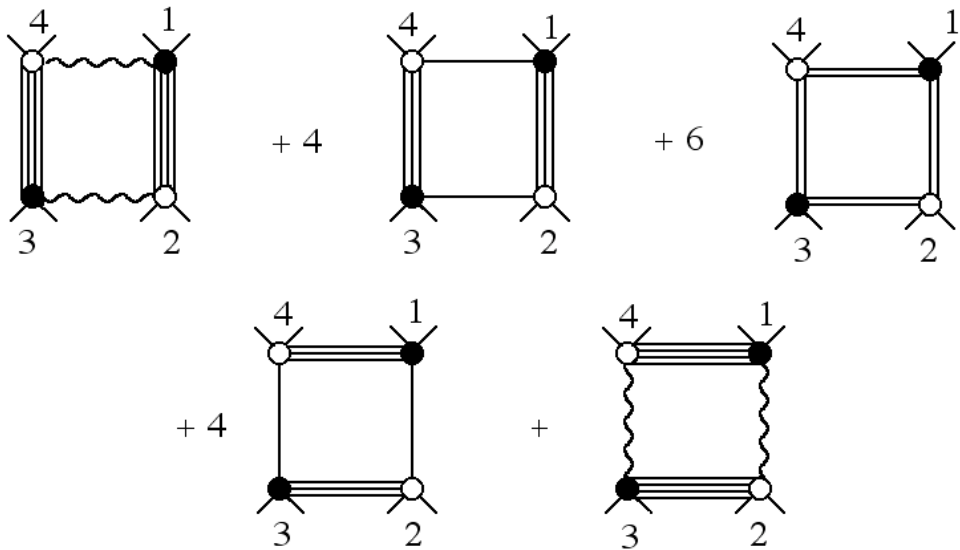

Note that the vital combinatorial factors 1, 4, 6, 4, 1 arise from the 'looping' property of the formal anticommuting numerators.

This sum corresponds, term by term, to the momentum-space expression:

$$\left( \frac{\langle 12\rangle^3 \langle 34\rangle^3}{\langle 23\rangle\langle 41\rangle\langle 13\rangle^4} - 4\frac{\langle 12\rangle^2 \langle 34\rangle^2}{\langle 13\rangle^4} + 6\frac{\langle 12\rangle\langle 34\rangle\langle 41\rangle\langle 23\rangle}{\langle 13\rangle^4} - 4\frac{\langle 41\rangle^2 \langle 23\rangle^2}{\langle 13\rangle^4} + \frac{\langle 41\rangle^3 \langle 23\rangle^3}{\langle 12\rangle\langle 34\rangle\langle 13\rangle^4} \right) \delta(\sum p^a)$$

$$= \left( \frac{(\langle 12\rangle\langle 34\rangle - \langle 41\rangle\langle 23\rangle)^4}{\langle 23\rangle\langle 41\rangle\langle 12\rangle\langle 34\rangle\langle 13\rangle^4} \right) \delta(\sum p^a)$$

$$= \left( \frac{\langle 24\rangle^4}{\langle 23\rangle\langle 41\rangle\langle 12\rangle\langle 34\rangle} \right) \delta(\sum p^a)$$

which is the correct amplitude.

Note that we have the 'square identity':

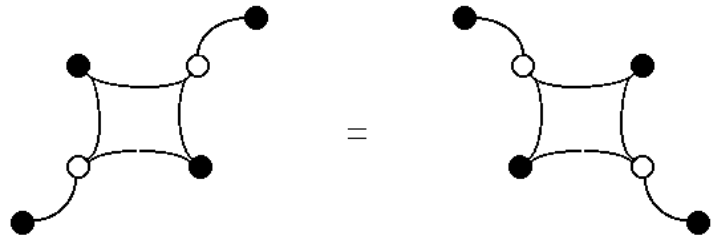

Making an assumption that two boundary lines connecting the same vertices are equivalent to a single boundary line, we can deduce the corollaries:

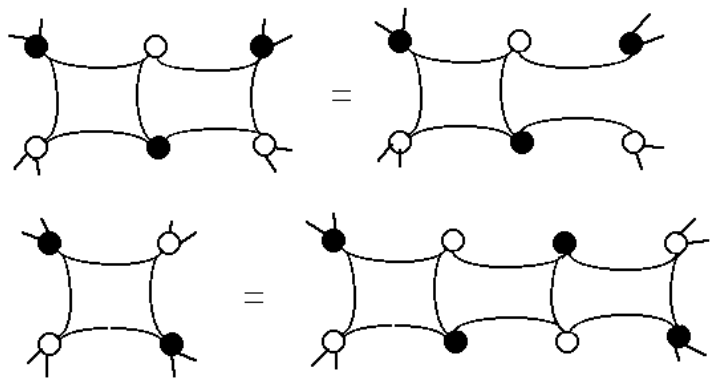

Identities of this type for double and triple-box diagrams have been known and used for long time in twistor diagram theory but have had to be examined case by case for different external homogeneities. The new notation unifies these results in a simply expressed form.

**5.2** A(12345)

In this case we obtain the sum of two diagrams. Suppose we pivot with field 2 on a dual twistor, field 3 on a twistor. We derive:

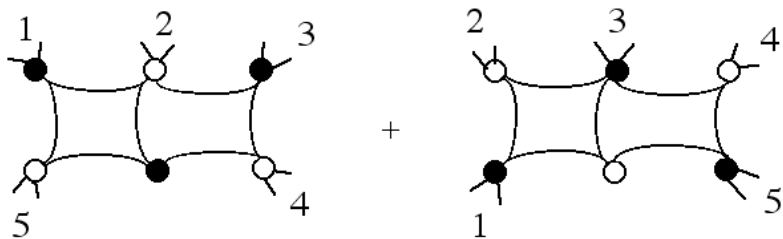

These two diagrams correspond to the (2 positive, 3 negative) and (3 positive, 2 negative) helicity sectors respectively. Note, however, that there is no need to separate out the helicity parts: each diagram simply vanishes outside its sector.

Other pivot choices will give other representations. But these representations can all be shown to be equivalent just by a graphical manipulation, using the square identity.

In the 'hard' case with the external fields all of homogeneity 0, the super-expansion defines a sum of fifteen terms. There is no point in writing out the individual terms; it follows directly from the form of the super-expansion coefficients that these fall into a trinomial pattern which is readily summed. This pattern extends to more general cases, and needs only elementary combinatorial algebra for evaluation.

## 5.3 A(12345678)

By repeated use of the generalised recursion rule, we can derive a sum of 42 terms, of which 20 are non-vanishing on the sector with four positive and four negative helicities. These 20 are given by the following diagrams. We have used the freedom in representation to put them in a form which makes their symmetries manifest: they are all presented as quiltings of an octagon by quadrilateral patches:

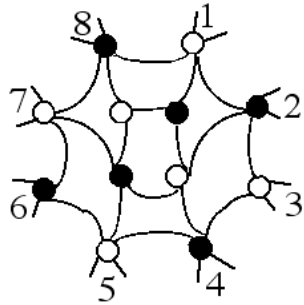

+ 7 similar terms

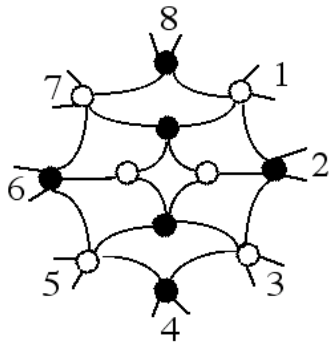

+ 3 similar terms

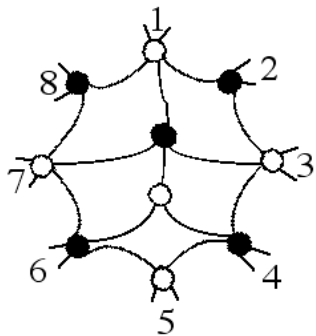

+ 7 similar terms

By 'similar terms' is meant, in each case, the diagrams obtained by keeping the exterior vertices and the exterior fields the same, but rotating the interior lines and dualizing the vertices as necessary.

There is a dual form of the amplitude in which the role of twistor and dual twistors is exchanged. The equality of the two summations implies a non-trivial 40-term linear identity between diagrams. It can be can be considered as an 'octagon identity' in analogy to the 'square identity' and the 6-term 'hexagon identity' which arises for A(123456).

We now restrict to cases where the external fields are helicity eigenstates.
In the case of A(+ − + − + − + −) the result is equivalent to the sum of 20 terms as given by Britto, Cachazo and Feng (2004b).

For instance the term

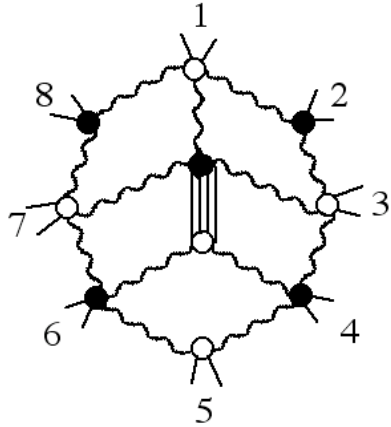

is equivalent to their expression $g^3 T$ :

$$\frac{[13]^4 [71]^4 \langle 46 \rangle^4}{[12][23][78][81]\langle 45\rangle\langle 56\rangle\langle 4|(2+3)|1]\langle 6|(7+8)|1][1|(2+3)(4+5+6)|7][3|(4+5+6)(7+8)|1]}$$

Roughly speaking, each internal line of the twistor diagram corresponds to a factor in the momentum-space expression. Having identified this correspondence, it can be applied to the other helicity patterns, which will differ in their arrangements of quadruple poles and boundary-lines, but will still involve the same basic factors. This saves one from starting afresh with each helicity assignment.

Besides this computational convenience, a deeper point about by these super-expanded diagrams is that they address the 'googly problem' as posed by Penrose: they yield a formalism in which it is possible to add the (–4) and 0-degree functions and so treat the external field as a unity.

## 6. Emergent features:

It is easy to calculate the number of terms that will arise through using the recursion rule. In the sector where there are $r$ positive fields and $(n - r)$ negative fields, the number is

$$\frac{(n-4)!(n-3)!}{(r-2)!(n-r-1)!(r-1)!(n-r-2)!}$$

The total number of terms for $n$ fields is

$$\frac{(2n-6)!}{(n-3)!(n-2)!}$$

which grows like $4^n$. Intuitively, one may say that introducing a further gluon multiplies the number of internal interaction possibilities by a factor of 4.

There is good reason to suppose that this is the smallest number of terms that can be achieved when using on-shell recursion for tree amplitudes with $n$ *general* fields. A general formula, of the kind illustrated above for 8 fields, may not be so useful when the fields are helicity eigenstates and have a special form, *e.g.* with many consecutive positive or negative helicities. Then it may well be more computationally efficient to use the various identities available to derive a different sum of 20 terms enjoying the property that many of them vanish for that particular helicity pattern.

Thus Roiban, Spradlin and Volovich (2004) gave a shorter 6-term formula for A(+ + + + – – – –). Their ingenious methods, featuring miraculous cancellations, are not in fact necessary. If one pivots on a (+ –) pair with the BCF formula (which was not available at the time of their work) it is readily seen that there are 14 vanishing contributions and that their 6-term formula emerges immediately.

This is in fact just a special case of a simplification that occurs for amplitudes with $r$ consecutive positive fields, $(n-r)$ consecutive negative fields. It is not hard to show that such an amplitude has a formula with just $\dfrac{(n-4)!}{(r-2)!(n-r-2)!}$ non-vanishing terms, a number which grows only like $2^n$. But this result does not indicate the existence of any new relations shortening the expressions in general.

Graphical methods may well give new ways of expressing, combining, and evaluating these terms. In particular, it would be useful to have a notation in which the square and higher identities are made implicit, so that the equivalence classes of quiltings of an *n*-agon are transparent.

In addition, it may be useful to bring the Ward identities into this graphical form, thus connecting diagrams with different colour-ordering. They too can be seen as linear relationships between super-expanded quilts.

We have only considered tree amplitudes. We leave open the question of loop Feynman amplitudes, which of course depend on the supersymmetric extension, if any. In general terms one would expect such amplitudes to correspond to twistor diagrams covering regions which are not simply connected. Similar general considerations also suggest specific forms that such diagrams might take. However, at present we can draw no definite conclusions. We cannot really make any statement about the extension of the twistor diagram theory without having a more general and satisfactory characterization of the procedure which has been defined above. This might be done either through a link to string theory, and hence to a more fundamental generating principle; or else through new developments in twistor geometry. Some observations pointing in the latter direction will be presented in a subsequent paper.

**Acknowledgments:**

The inspiration of Roger Penrose, who first wrote down twistor diagrams more than thirty years ago, is of paramount importance. The more recent developments in twistor diagram theory have been greatly stimulated by the twistor string theory of Edward Witten (2003), and by the many distinguished contributors to the January 2005 'Twistors and Strings' workshop at Oxford University and the November 2005 'From Twistors to Amplitudes' workshop at Queen Mary, London University. I am also grateful to Philip Candelas for a specific suggestion, and to David Skinner for much helpful and stimulating discussion as well as the particular application of super-algebra noted in §3.

Version 1 of this paper was submitted on 29 December 2005. In Version 2 of this paper, dated 11 January 2006, I have corrected some typographical mistakes, made small improvements in wording and introduced another citation.
Further supporting material will appear on www.twistordiagrams.org.uk